\documentclass{article}
\usepackage[dvips]{graphicx}

\usepackage{proof}
\usepackage{stmaryrd}
\usepackage{smatsu}
\usepackage{latexsym}
\usepackage{times}
\usepackage[T1]{fontenc}
\normalfont
\usepackage{mathptm}

\begin{document}
\title{Nondeterministic Linear Logic}
%\title{Nondeterministic Linear Logic\\
%  (Extended Abstract)}
\author{Satoshi Matsuoka\thanks{Department of Electrical and Computer
    Engineering, Faculty of Engineering, Nagoya Institute of Technology,
    Gokiso, Showa-Ku, 466, Japan.}\\
  {\bf {\tt matsuoka@juno.ics.nitech.ac.jp}}}
\date{}
\maketitle
\begin{abstract}
In this paper, we introduce Linear Logic with a nondeterministic facility, 
which has a self-dual additive connective. 
In the system the proof net technology is available in a natural way.
The important point is that nondeterminism in the system is expressed by 
the process of normalization, not by proof search. 
Moreover we can incorporate the system into Light Linear Logic and 
Elementary Linear Logic developed by J.-Y.Girard recently: 
Nondeterministic Light Linear Logic and Nondeterministic 
Elementary Linear Logic are defined in a very natural way.
\end{abstract}
\newtheorem{lemma}{Lemma}
\newtheorem{theorem}{Theorem}
\newtheorem{definition}{Definition}
\newtheorem{corollary}{Corollary}
\newtheorem{proposition}{Proposition}
\newtheorem{remark}{remark}
\newenvironment{proof}{\begin{flushleft}{\it Proof.} \ \ }{\end{flushleft}}

\def\labelitemi{$\bullet$}
\def\labelitemii{$\star$}
\section{Introduction}
So far (untyped or typed) lambda calculi with the facility of 
nondeterminism have been studied:
recently e.g., in \cite{Aba94,DCLP93}. 
%However such calculi are only lambda calculi with the facility of
%nondeterminism.
For example, in \cite{DCLP93} nondeterminism is represented by using union
type, while parallelism by using intersection type: 
this means that nondeterminism corresponds to the logical connective ``or'' 
and parallelism to ``and''. 
Further this means that nondeterminism and parallelism are dual notions each
other.
Basically other researchers similarly classify nondeterminism and parallelism.
In this paper, we advocate that nondeterminism and parallelism are not dual
notions. For this we use the framework of Linear Logic \cite{Gir87}. 
In Linear Logic, usual logical connectives are classified into two: 
multiplicative and additive connectives. 
Our advocacy is that nondeterminism and parallelism are classified in 
``computation as normalization'' paradigm as follows:
\begin{itemize}
\item Nondeterminism = Additive.
\item Parallelism = Multiplicative.
\end{itemize}
Already it has been pointed out that 
the multiplicative connectives are deeply related to parallelism since 
the appearance of \cite{Gir87}. Currently V.Pratt studies the relationship 
intensively in the context of Chu space \cite{Pra95,Pra97}. 
Here we point out that
the additive connectives are deeply related to nondeterminism. 
We incorporate nondeterminism facility into the framework of 
Linear Logic by introducing new additive connective $\NDWITH$ 
(nondeterministic with), which is self-dual. 
In the framework, nondeterminism is 
represented by reduction of cut between two $\NDWITH$: 
by the reduction of $\NDWITH$ from one proof net two proof nets are
obtained.
Note that standard Linear Logic is deterministic: this means 
proof nets have Church-Rosser property in the syntactical level and 
there are some  ``deterministic'' denotational semantics of Linear Logic 
in the semantical level. So in order to incorporate nondeterminism into 
the framework of Linear Logic, we must introduce the new connective.
%By using $\NDWITH$ we can define Nondeterministic Light Linear Logic and
%Nondeterministic Elementary Linear Logic in a very natural way. 
Our advocacy has not been advocated before as far as we know in the context
of ``computation as normalization'' paradigm.  
Also I believe that such a classification contributes to studies w.r.t. 
relationship between Linear Logic and Process Calculus. \\
On the other hand, our Nondeterministic Linear Logic also contributes to 
the study of the logical aspect of Complexity Theory in the context of 
``computation as normalization'' paradigm. 
We can encode any nondeterministic polynomial-time Turing Machine into 
a proof (or a proof net) of Nondeterministic Light Linear Logic. 
The encoding is 
a nondeterministic version of Girard's encoding deterministic
polynomial-time Turing Machines into proofs of Light Linear Logic. 
But our polymorphic encoding of nondeterministic computations is original. 
Also by using the same method, we can formulate Nondeterministic Elementary 
Linear Logic and prove the similar statement. 
This is the main technical contribution of this paper.
\section{The System}
\label{SYSNDMALL}
The system NDMALL is usual MALL (the multiplicative additive fragment of 
Linear Logic) with  $\NDWITH$ (nondeterministic with). 
The connective has arity 2 (hence in NDMALL $A \NDWITH B$ is accepted as 
a formula if $A$ and $B$ are NDMALL formulas). 
The negation of $A \NDWITH B$ is defined as follows: 
\[ (A \NDWITH B)^\bot \equiv_{\mathdef} A^\bot \NDWITH B^\bot \]
The inference rules for NDMALL are the same as MALL except for 
the following rule:
\[ \infer[(\rm{NDWITH})]{\vdash \Gamma, A \NDWITH B}
{\vdash \Gamma, A & \vdash \Gamma, B} \]
The notion of proofs (in sequent calculus) of NDMALL is defined in usual
manner. 
Obviously the connective $\NDWITH$ belongs to additives.
The problematic point of NDMALL is that
any sequent of the form $\vdash A \NDWITH B, A^\bot \NDWITH B^\bot$ does not 
has the proof of $\eta$-long normal form, i.e., consisting of just 
atomic formulas. But for example, many modal logics also do not 
have such proofs, even cut-free systems. We believe Nondeterministic Linear
Logic can be accepted as a logical system. Though even you does not agree with 
the belief, you should accept our system as a type system for
nondeterministic computations.\\
In practice, the connective does not occur in conclusions of NDMALL proofs: 
if it occurs in them, then in one sided sequent calculus (or in the
formulation of proof nets) it behaves like $\WITH$ in completely the same
manner. Hence we can assume that $\NDWITH$ does not occur in cut free NDMALL 
proofs.  
We omit cut elimination procedure for $\NDWITH$ in NDMALL sequent calculus. 
But we will introduce the procedure using NDMALL proof nets 
in Section~\ref{SimpleOS}.
\section{NDMALL proof nets}
First we shall define NDMALL proof structures, which are basically 
the same as them in \cite{Gir95b} except for connective $\NDWITH$. 
Simply by formulas we mean NDMALL formulas. Note that 
to each $\NDWITH$-link $L$ an eigenweight $p_L$ is assigned.
\begin{definition}\label{deflink}
A link $L$ is an $n+m$-tuple of formulas with a type:
$\displaystyle
\frac{P_1,\ldots,Pn}{Q_1,\ldots,Q_m} L
$
The type of a link is either ID, Cut, generalized axiom, $\TENS$, 
$\PAR$, $\WITH$, $\PLUS_1$, $\PLUS_2$, or $\NDWITH$.
To each type, $n$, a number of its premises and $m$, a number of its 
conclusions are assigned($m,n \ge 0, \, m+n \neq 0$). 
The links with ID, Cut, generalized axiom, $\TENS, \PAR, \WITH, \PLUS_1$,
$\PLUS_2$, and $\NDWITH$ 
as types have the following forms:
\begin{center}
%  {\bf Identity Links:}
  \[ \mbox{ID-links} \, \, \,  \frac{}{A \quad A^\bot} \quad
     {\rm Cut \, \, \, links} \, \, \,  \frac{A \quad A^\bot}{\rm Cut}
\quad
\mbox{generalized axiom-links} \, \, \, \frac{}{A_1 \cdots A_n}
\]
%  {\bf Multiplicative Links:}
  \[
  {\rm times}  \, \, \, \frac{A \quad B}{A \TENS B}
  \quad
  {\rm par}  \, \, \, \frac{A \quad B}{A \PAR B}
\quad
%  {\bf Additive Links:}
  {\rm with}  \, \, \, \frac{A \quad B}{A \WITH B} \quad
  {\rm plus} \quad
  \PLUS_1 \, \,  \frac{A}{A \PLUS B}
  \quad
  \PLUS_2 \, \, \frac{B}{A \PLUS B} 
\]
\[
  \mbox{nondeterministic with}  \, \, \, \frac{A \quad B}{A \NDWITH B}
  \]
\end{center}
We must distinguish a left premise ($A$) and a right premise ($B$) in 
$\TENS, \PAR$ and $\WITH$-links. For example, in a $\NDWITH$-link with 
$A \NDWITH A$ as the conclusion, the two premises $A$ and $A$ must be
distinguished in an obvious way.
\end{definition}

\begin{definition}
    To any $\WITH$-link or $\NDWITH$-link 
    $L$ with $A \WITH B$  or $A \NDWITH B$ as its conclusion, 
    we associate an eigenweight $p_L$, which is a boolean variable. 
    The intuitive meaning of $p_L$ is the choice $\{ l/r \}$ between the premises
    $A$ and $B$: $+ p_L$ stands for the selection ``left'', i.e., $A$ and 
    $- p_L$ stands for the selection ``right'', i.e., $B$.
    We use $\epsilon . p_L$ to speak
    of $+ p_L$ or $- p_L$.
\end{definition}
\begin{definition}
\label{proofstr}
A triple $\Theta=(V,E,w)$ is a proof structure if
\begin{itemize}
\item $(V,E)$ is a pair such that 
$V$ is a multiset of formulas and $E$ is 
a multiset of links between formulas occurring in $V$. 
\item $w$ is a function such that\\
(i) For each formula $A$ in $V$, a weight $w(A)$, i.e.,
  a non-zero element of the boolean algebra generated by the eigenweights 
  $p_1,\ldots,p_n$ of the $\WITH$-links or $\NDWITH$-links of $\Theta$;\\
(ii) For each link $L$ in $E$, a weight $w(L)$, i.e.,
  a non-zero element of the boolean algebra generated by the eigenweights 
  $p_1,\ldots,p_n$ of the $\WITH$-links or $\NDWITH$-links of $\Theta$.
\end{itemize}
Moreover, the following conditions must be satisfied:
\begin{enumerate}
\item [(a)] Each formula in $V$ is the premise of at most one link and the
  conclusion of at least one link. 
%  So each formula may become the  conclusion of 
%  several links unlike in multiplicative proof structures; 
  The formulas which
  are not premises of some link are called the conclusions of $\Theta$;
\item [(b)] $\displaystyle w(A) = \sum_{L \, \mbox{\scriptsize has} \, A \, 
    \mbox{\scriptsize as the conclusion}} w(L)$; 
\item [(c)] If $A$ is a conclusion of $\Theta$, then $w(A)=1$;
\item [(d)] If $u$ is any weight occurring in $\Theta$, then $u$ is a monomial 
  $\epsilon_1 . p_{L_1} \cdots \epsilon_n . p_{L_n}$ of 
  eigenweights and negations of eigenweights;
\item [(e)] If $u$ is a weight occurring in $\Theta$ and containing 
  $\epsilon . p_L$ then $u \le w(L)$;
\item [(f)] If $L$ is any non ID-link, with premises
  $A$ and/or $B$ then
\begin{itemize}
\item if $L$ is any of $\TENS, \PAR$ and Cut, then 
  $w(L)=w(A)=w(B)$;
\item if $L$ is a $\PLUS_1$-link, then $w(L)=w(A)$;
\item if $L$ is a $\PLUS_2$-link, then $w(L)=w(B)$;
\item if $L$ is a $\WITH$-link, then $w(A)=w(L) \cdot p_L$ and 
  $w(B) = w(L) \cdot \neg p_L$;
% \ (hence $w(L)=w(A)+w(B)$);
\item if $L$ is a $\NDWITH$-link, then $w(A)=w(L) \cdot p_L$ and 
  $w(B) = w(L) \cdot \neg p_L$;
% \ (hence $w(L)=w(A)+w(B)$).
\end{itemize}
\item [(g)] For any $A \in V$, if the links whose conclusion is $A$ are 
$L_1,\ldots,L_m$ then for each $1 \le i,j \le m$, whenever $i \neq j$, then
$w(L_i) \neq w(L_j)$.
\end{enumerate}
\end{definition}

\begin{definition}
  Let $\phi$ be a valuation for a proof structure $\Theta=(V,E,w)$, 
  i.e. a function from the set of
  eigenweights of $\Theta$ to $\{ 0,1 \}$, which is extended to a function
  (still denoted $\phi$) from the weights of $\Theta$ to $\{ 0,1 \}$. 
  A pair $\phi(\Theta)=(V_0, E_0)$ is the slice by $\phi$ if 
  $V_0$ is the restriction to the formulas $A$ in $V$ such that $\phi(w(A))=1$
  and $E_0$ is the restriction of $E$ by $V_0$ where 
  the definition of $\WITH$-links and $\NDWITH$-links is changed such that 
  they have exactly one premise and one conclusion.
\end{definition}

The definition of the dependencies of the weights and the formulas in proof
structures on an eigenweight is the same as that of \cite{Gir95b}. 

\begin{definition}
  Let $\phi$ be a valuation of $\Theta$, let $p_L$ be an eigenweight.
  A weight $w$ (in $\Theta$) depends on $p_L$ 
  (in $\phi(\Theta)$) if $\phi(w) \neq \phi_L(w)$, where the valuation 
  $\phi_L$ is defined as follows:
  \begin{itemize}
  \item $\phi_L(p_L) = \neg (\phi(p_L));$
  \item $\phi_L(p_{L'}) = \phi(p_{L'})$ if $L' \neq L$.
  \end{itemize}
  A formula $A$ of $\Theta$ is said to depend on $p_L$ (in $\phi(\Theta)$),
  if $A$ is the conclusion of a link $L'$ such that $\phi(w(L'))=1$ and 
  $\phi_L(w(L'))=0$.
\end{definition}
\def\select{\mathop {\rm select} \nolimits}
\begin{definition}
    A switching 
    ${\cal S}=(\phi_{\cal S}, \select_\PAR, \select_\WITH, \select_\NDWITH)$ 
    of a proof structure $\Theta$ consists in:
\begin{itemize}
\item A choice of a valuation $\phi_{\cal S}$ for $\Theta$;
\item A function $\select_\PAR$ from the set of all $\PAR$-links $L$ of 
$\phi_{\cal S}(\Theta)$ to $\{l,r\}$ whose element represents a choice for
premises of a $\PAR$-link.
\item A selection $\select_\WITH$
  for each $\WITH$-link $L$ of $\phi_{\cal S}(\Theta)$ 
  a formula $\select_\WITH(L)$, the jump of $L$, depending on $p_L$ in 
  $\phi_{\cal S}(\Theta)$. There is always a normal choice of jump for $L$,
  namely the premise $A$ of $L$ such that $\phi_{\cal S}(w(A))=1$.
\item A selection $\select_\NDWITH$
  for each $\NDWITH$-link $L$ of $\phi_{\cal S}(\Theta)$ 
  a formula $\select_\NDWITH(L)$, the jump of $L$, depending on $p_L$ in 
  $\phi_{\cal S}(\Theta)$. There is always a normal choice of jump for $L$,
  namely the premise $A$ of $L$ such that $\phi_{\cal S}(w(A))=1$.
\end{itemize}
\end{definition}

\begin{definition}
    Let ${\cal S}$ be a switching of a proof structure $\Theta$;\\
    the graph $\Theta_{\cal S} = (V_{\cal S}, E_{\cal S})$ corresponding to 
    ${\cal S}$ consists in:
    \begin{itemize}
    \item the vertices $V_{\cal S}$ is 
      $V_0$ of $\phi_{\cal S}(\Theta) = (V_0,E_0)$;
    \item the edges $E_{\cal S}$ are consists of:
    \begin{enumerate}
    \item the edge between the conclusions for any 
      $ID$-link of $\phi_{\cal S}(\Theta)$;
    \item the edge between the premises for any
      Cut-link of $\phi_{\cal S}(\Theta)$;
    \item the edge between the conclusion and the premise for any
      $\PLUS$-links of $\phi_{\cal S}(\Theta)$;
    \item  the edges between the left premise and the conclusion, and between 
      the right premise and the conclusion for any 
      $\TENS$-link of $\phi_{\cal S}(\Theta)$;
    \item the edge between the 
      the premise (left or right) selected by $\select_\PAR(L)$ and the
      conclusion of any $\PAR$-links $L$ of $\phi_{\cal S}(\Theta)$;
    \item the edge between the jump $\select_\WITH(L)$ 
      of $L$ and the conclusion for any $\WITH$-link L.
    \item the edge between the jump $\select_\NDWITH(L)$ 
      of $L$ and the conclusion for any $\NDWITH$-link L.
    \end{enumerate}
\end{itemize}
\end{definition}

\begin{definition}
    A proof structure $\Theta$ is said to be a proof net if for any
    switching ${\cal S}$, the graph $\Theta_{\cal S}$ is connected and
    acyclic. 
\end{definition}

The removal of a link of a proof structure $\Theta$ in NDMALL 
is defined in the same manner as \cite{Gir95b} except for $\NDWITH$-links. 
Here the definition of the removal for $\NDWITH$-links is only added.
\begin{definition}
\begin{itemize}
\item The case where 
  $L$ is a $\NDWITH$-link with premises $A$ and $B$ such that 
  $w(L)$=1 and $L$ is a conclusion of $\Theta$, and 
  $\Gamma, A \NDWITH B$ is the set of conclusions of $\Theta$. 
  The removal of $L$ is the operation which first removes 
  the conclusion $A \NDWITH B$
  and the link $L$, gets a proof structure $\Theta'$ 
  and then forms two proof structures $\Theta_A$ and $\Theta_B$ from $\Theta'$:
\begin{itemize}
\item In $\Theta'$ make the substitution $p_L=1$, and keep only those links 
  $L'$ whose weight is still non-zero, together with the premises and
  conclusions of such links: the result is by definition $\Theta_A$, a proof
  structure with conclusions $\Gamma, A$.
\item In $\Theta'$ make the substitution $p_L=0$, and keep only those links 
  $L'$ whose weight is still non-zero, together with the premises and
  conclusions of such links: the result is by definition $\Theta_B$, a proof
  structure with conclusions $\Gamma, B$.
\end{itemize}
\end{itemize}
\end{definition}
\begin{definition}
    A proof structure $\Theta$ is sequentializable if 
    \begin{enumerate}
    \item $\Theta$ is an ID-link, or;
    \item the proof structures which are obtained by the removal of 
      a terminal link in $\Theta$ are sequentializable.
    \end{enumerate}
%    it can be reduced, by
%    iterated removal of terminal links, to identity links.
\end{definition}

The proof of the following theorem is completely the same as that of 
\cite{Gir95b} which uses the empire for each valuation and each formula, since 
in fixed proof nets $\NDWITH$-links behave in the same manner as
$\WITH$-links. 
However the behavior of $\NDWITH$-link in cut elimination is different from 
that of $\WITH$ which is defined in the next section.

\begin{theorem}[\cite{Gir95b}]
  $\Theta$ is a proof net iff $\Theta$ is sequentializable.
\end{theorem}

\section{Lazy Cut Elimination in NDMALL}\label{SimpleOS}
\begin{definition}
  A cut-link $L$ is ready if 
\begin{itemize}
\item $w(L) = 1$ and;
\item If the premises of $L$ are $A$ and $A^\bot$ then 
  both $A$ and $A^\bot$ are the conclusion of exactly one link.
\end{itemize}
\end{definition}

\begin{definition}[lazy cut elimination]
Let $L_0$ be a ready cut in a proof net $\Theta$, whose premises 
$B \NDWITH C$ and $B^\bot \NDWITH C^\bot$ are the respective conclusions of 
links $L$ and $L'$. Then we define the contractums $\Theta'$ and $\Theta''$ 
of redex $\Theta$
when reducing $L_0$ in $\Theta$. 
\begin{itemize}  
\item If $L$ is a $\NDWITH$-link (with premises $B$ and $C$) and 
$L'$ is a $\NDWITH$-link (with premises $B^\bot$ and $C^\bot$), then 
$\Theta'$ and $\Theta''$ are obtained in three steps 
(the reduction is called $\NDWITH$-reduction):\\
  {\bf how to get $\Theta'$ (resp. $\Theta''$):}\\
  First we remove in $\Theta$ the formulas 
  $B \NDWITH C$ and $B^\bot \NDWITH C^\bot$ as well as $L_0$, $L$ and $L'$; 
  then we replace the eigenweights $p_L$ and $p_L'$ by $1$ (resp. $0$) and keep
  only 
  those formulas and links that still have a nonzero weight: therefore 
  $B$(resp. $C$) and $B^\bot$ (resp. $C^\bot$) 
  remain with weight $1$ whereas $C$ (resp. $B$) and $C^\bot$ (resp. $B^\bot$) 
  disappears; finally we add a cut between $B$ (resp. $C$) and $B^\bot$ 
  (resp $C^\bot$), and then 
  get $\Theta'$ (resp. $\Theta''$).
\end{itemize}
\end{definition}
\begin{proposition}
  If $\Theta'$ is obtained from a proof net $\Theta$ by lazy cut
  elimination, then $\Theta'$ is a proof net and has the same conclusions as 
  $\Theta$.
\end{proposition}
\begin{proof}
We only consider $\NDWITH$-reduction:
\begin{itemize}
\item Here we use the same meta symbols as the definition of lazy cut
  elimination. \\
  \begin{enumerate}
  \item to show $\Theta'$ is a proof net:
    Let $\phi'$ be a valuation for $\Theta'$. 
    Then we define the valuation $\phi$ for $\Theta$ from $\phi'$ as follows:
    \[ 
    \phi(p_M) = \cases {
      1 & \mbox{if} \, $M = L$ \, \mbox{or} \, L' ; \cr
      \phi'(p_M) & \mbox{if otherwise}. \cr
      }
    \]
    Let ${\cal S'}$ be any switching with the valuation $\phi'$ for $\Theta'$.
    Then we define the switching ${\cal S}$ for $\Theta$ from ${\cal S'}$ as
    follows: 
    \begin{itemize}
    \item the valuation of ${\cal S}$ is $\phi$;
    \item $\select^{\cal S}_\PAR$ and $\select^{\cal S}_\WITH$ 
      are the same as ${\cal S'}$;
    \item $\select^{\cal S}_\NDWITH(M) = \cases {
        B & \mbox{if} \, $M = L$; \cr
        B^\bot  & \mbox{if} \, $M = L'$ ; \cr
        \select^{\cal S'}_\NDWITH(M) & \mbox{if otherwise}. \cr
        }$
    \end{itemize}
    Since $\Theta$ is a proof net by assumption, 
    $\Theta_{\cal S}$ is acyclic and connected. 
    Then from this it is immediate that $\Theta'_{\cal S'}$ is acyclic and
    connected. Hence $\Theta'$ is a proof net.
  \item to show $\Theta''$ is a proof net:
    Let $\phi''$ be a valuation for $\Theta''$. 
    Then we define the valuation $\phi$ for $\Theta$ from $\phi''$ as follows:
    \[ 
    \phi(p_M) = \cases {
      0 & \mbox{if} \, $M = L$ \, \mbox{or} \, L'; \cr
      \phi''(p_M) & \mbox{if otherwise}. \cr
      }
    \]
    Let ${\cal S''}$ be any switching for $\Theta''$ with the valuation 
    $\phi''$.
    Then we define the switching ${\cal S}$ for $\Theta$ from ${\cal S''}$ as
    follows: 
    \begin{itemize}
    \item the valuation of ${\cal S}$ is $\phi$;
    \item $\select^{\cal S}_\PAR$ and $\select^{\cal S}_\WITH$ 
      are the same as ${\cal S''}$;
    \item $\select^{\cal S}_\NDWITH(M) = \cases {
        C & \mbox{if} \, $M = L$; \cr
        C^\bot  & \mbox{if} \, $M = L'$; \cr
        \select^{\cal S''}_\NDWITH(M) & \mbox{if otherwise}. \cr
        }$
    \end{itemize}
    Since $\Theta$ is a proof net by assumption, 
    $\Theta_{\cal S}$ is acyclic and connected. 
    Then from this it is immediate that $\Theta''_{\cal S''}$ is acyclic and
    connected. Hence $\Theta''$ is a proof net. $\Box$
  \end{enumerate}
\end{itemize}
%  \qed
%$\Box$
\end{proof}
Since $\NDWITH$ is a variant of additive connectives, by the same method 
as that in \cite{Gir95b}, the following proposition is easily proved.
\begin{proposition}
\label{LAZYMALL}
  By lazy cut elimination, any MALL proof net is reduced to a unique 
  normal form (which contains ready cuts) in linear time of its size.
\end{proposition}

%\section{Programming with NLL}
\section{Nondeterministic Light Linear Logic}
\newcommand{\NEUT}
{\displaystyle \S}

In \cite{Gir95c}, it is shown that 
(1) any p-time Deterministic Turing Machine are representable in 
Light Linear Logic (for short LLL) and 
(2) under the condition of bounded depth any LLL proof net is reduced to
a normal form in p-time of its size. 
In this section we show that 
(1') any p-time Nondeterministic Turing Machine are representable in 
Nondeterministic Light Linear Logic (for short NDLLL) and 
(2') under the condition of bounded depth any NDLLL proof net is reduced to
a normal form by lazy cut elimination in p-time of its size. 
The system NDLLL is obtained from LLL by adding the inference rule 
(NDWITH) in Section~\ref{SYSNDMALL}. 
It is not difficult to show (2') if we follow Girard's proof for LLL, since 
any NDMALL proof net is reduced 
a normal form by lazy cut elimination in linear time of its size 
(Proposition~\ref{LAZYMALL}) and 
$\NDWITH$ connective does not interact with any exponential connectives. \\
Let a Nondeterministic Turing Machine be $M$. 
Let $\Sigma$ be the set of the symbols used in $M$ and 
${\cal Q}$ be the set of the states used in $M$. 
Let $p$ be the number of the symbols used in $M$, i.e., the cardinal of
$\Sigma$ and $q$ be the number of the
states used in $M$, i.e., the cardinal of ${\cal Q}$. 
In order to prove (1'), we only show the move (transition) relation of 
the Nondeterministic Turing Machine $M$ is representable in NDMALL, since 
from a representation in NDMALL of the move relation of $M$ we can easily 
construct 
a proof net with $!^k 1 \TENS {\bf Tur}^{p,q} \LIMP {\bf Tur}^{p,q}$ 
as the conclusion that represents $M$
completely, where 
${\bf Tur}^{p,q} = {\bf list}^p \TENS {\bf list}^q \TENS {\bf bool}^q$, 
${\bf list}^p = 
\forall X. (\overbrace{!(X \LIMP X) \LIMP \cdots !(X \LIMP X)}^p) \LIMP 
\NEUT (X \LIMP X)$, and 
${\bf bool}^k = \forall X. \NEUT 
(\overbrace{X \WITH \cdots \WITH X}^k \LIMP X)$.

The move relation $R$ of $M$ is represented as a subset of 
$(\Sigma \times {\cal Q}) \times (\Sigma \times {\cal Q} \times 
\{\leftarrow,\rightarrow\})$.
Then it is sufficient to represent the move relation $R$ 
by a NDLLL proof net with 
${\bf bool}^{p \times q}
\LIMP {\bf bool}^{p \times q \times 2}$ as the
conclusion, 
since
we can easily see the set $(\Sigma \times {\cal Q})$ is represented
by ${\bf bool}^{p \times q}$ and 
$(\Sigma \times {\cal Q} \times 
\{\leftarrow,\rightarrow\})$ 
by ${\bf bool}^{p \times q \times 2}$, 
we can easily construct any proof net with 
${\bf bool}^p \TENS {\bf bool}^q \LIMP {\bf bool}^{p \times q}$ as 
the conclusion and with 
${\bf bool}^{p \times q \times 2} \LIMP 
{\bf bool}^p \TENS {\bf bool}^q \TENS {\bf bool}^2$ as the conclusion 
by using a general version of ${\sf D}$ in Section 11.3 in \cite{GLT89}, 
and given any proof net with
${\bf bool}^{p \times q}
\LIMP {\bf bool}^{p \times q \times 2}$ as the
conclusion, 
by composing these proof nets 
we can easily construct 
any proof net with 
${\bf bool}^p \TENS {\bf bool}^q \LIMP 
{\bf bool}^p \TENS {\bf bool}^q \TENS {\bf bool}^2$ as the conclusion.
%and finally we can construct the proof nets representing 
%the transitions of Nondeterministic Turing Machines. 
Let $m$ be 
$\max \{ |\{ (y,t,d) : (x,s,(y,t,d)) \in R \}| : x \in \Sigma, s \in {\cal
  Q}\}$. 
The following NDLL proof corresponds to the intended proof net:
\[
\infer={
\infer[(\NEUT)]{
\infer[(\EXISTS)]
{
\infer[(\ALL)]
{\infer[(\LIMP)]{\vdash {\bf bool}^{p \times q}\LIMP {\bf bool}^{p \times q \times 2}}
{{\bf bool}^{p \times q} \vdash {\bf bool}^{p \times q \times 2}}}
{{\bf bool}^{p \times q} \vdash 
\NEUT (\overbrace{X \WITH \cdots \WITH X}^{p \times q \times 2} \LIMP X)}
}
{
\NEUT(\overbrace{(\overbrace{X \NDWITH \cdots \NDWITH X}^{m}) \WITH \cdots 
\WITH 
(\overbrace{X \NDWITH \cdots \NDWITH X}^{m})}^{p \times q} \LIMP 
(\overbrace{X \NDWITH \cdots \NDWITH X}^{m}))
\vdash 
\NEUT (\overbrace{X \WITH \cdots \WITH X}^{p \times q \times 2} \LIMP X)}}
{\overbrace{(\overbrace{X \NDWITH \cdots \NDWITH X}^{m}) \WITH \cdots 
\WITH 
(\overbrace{X \NDWITH \cdots \NDWITH X}^{m})}^{p \times q} \LIMP 
(\overbrace{X \NDWITH \cdots \NDWITH X}^{m})
\vdash 
\overbrace{X \WITH \cdots \WITH X}^{p \times q \times 2} \LIMP X}}
{\infer=[(\WITH)] {\overbrace{X \WITH \cdots \WITH X}^{p \times q \times 2} 
\vdash
\overbrace{(X \NDWITH \cdots \NDWITH X) \WITH \cdots \WITH (X \NDWITH \cdots \NDWITH
  X)}^{p \times q}}
{
\infer*{\overbrace{X \WITH \cdots \WITH X}^{p \times q \times 2} \vdash X \NDWITH \cdots \NDWITH X}{\langle 1 \rangle} & \infer*{\cdots}{\cdots} & 
\infer*{\overbrace{X \WITH \cdots \WITH X}^{p \times q \times 2} \vdash X \NDWITH \cdots \NDWITH X}{\langle p \times q \rangle}}
& 
\infer=[(\NDWITH)]{\overbrace{X \NDWITH \cdots \NDWITH X}^m \vdash X}
{X \vdash X & \cdots & X \vdash X}
}
\]
The programming of the move relation $R$ corresponds to
the proofs between $\langle 1 \rangle$ and $\langle p \times q \rangle$, 
the move relation $R$. 
By the way there may be $x \in \Sigma$ and $s \in {\cal Q}$ such that
$|\{ (y,t,d) : (x,s,(y,t,d)) \in R \}| < m$. Then 
we introduce a new state ``halt'' and can construct a new proof net with 
$\vdash {\bf bool}^{p \times q} \LIMP {\bf bool}^{p \times (q+1) \times 2}$ 
as the conclusion from 
the already obtained proof net by turning 
$m - |\{ (y,t,d) : (x,s,(y,t,d)) \in R \}|$ transitions 
into ``halt'' state. 

From what precedes the following theorem is proved. 
%In the following, we shall present a representation of the function in
%NDLLL. 
% NDTM $B$N(B move function $B$O(B p$B!_(Bq $B$+$i(B p$B!_(Bq$B!_(B2 $B$X$NHs7hDjE*4X?t$H$7$F(B
% $BI=8=$G$-$k!#(B
\begin{theorem}
Any p-time Nondeterministic Turing Machine are representable in 
Nondeterministic Light Linear Logic.
\end{theorem}
It is obvious that in the context of Elementary Linear Logic, the same 
theorem is proved.

\section{Concluding-remarks}
As to the semantics of Nondeterministic Linear Logic, in this paper, 
we just presented a very primitive operational semantics: 
lazy cut elimination procedure (see Section~\ref{SimpleOS}). 
In order to justify Nondeterministic Linear Logic, 
we must develop denotational and operational semantics 
of Nondeterministic Linear Logic in more sophisticated ways:
\begin{itemize}
\item Denotational Semantics\\
In the usual coherent semantics,  self-dual connectives like $\NDWITH$
connective are not allowed. However in \cite{Gir96}, 
J.-Y.Girard has developed a semantics 
not only accommodating usual connectives of Linear logic, but also 
self-dual additive connectives: in the semantics formulas are interpreted by 
coherent Banach spaces (which are named by Girard) and 
proofs by vectors in the spaces. Since in NDMALL the Church-Rosser property
does not hold and the result of normalization of a proof leads many normal 
proofs, the NDMALL proofs in the semantics are interpreted by 
the sum of some vectors (i.e., a vector) in the coherent Banach spaces. 
The details will be left elsewhere. \\
Also interpretations of NDLL into Chu spaces \cite{Pra95,Pra97} and 
Game Semantics \cite{AG94} are interesting. Such researches will lead some insights 
on the relationship between Linear Logic and Concurrency Theory. 
\item Operational Semantics\\
For Linear Logic very elegant operational semantics have been developed: 
Geometry of Interaction (for short GOI). In \cite{Gir95a}, 
GOI has been extended to the system accommodating the additives.
The study of GOI for 
Nondeterministic Linear Logic is interesting. 
%In GOI for MALL a logic programming language with complex coefficients 
%is used. 
In GOI for MALL a simple logic programming language is used. 
It is not difficult to incorporate nondeterminism with logic
programming. Hence it seem that the development of GOI for NDMALL is not
so difficult. 
\end{itemize} 
%On the other hand, it seems possible that
%the method in this paper can be applied to 
%interaction combinators in \cite{Laf95} in order to 
%add the nondeterministic facility to it.


\begin{thebibliography}{1}
\bibitem[Aba94]{Aba94}
M. Abadi.
\newblock A Semantics for Static Type Inference in a Nondeterministic
Language. 
\newblock {\it Information and Computation}, 109:300--306, 1994.

\bibitem[AG94]{AG94}
S. Abramsky and R. Jagadeesan.
\newblock Games and full completeness for multiplicative linear logic.
\newblock JSL 59, pages 543-574, 1994.

\bibitem[DCLP93]{DCLP93}
M. Dezani-Ciancaglini, U. de'Liguoro and A. Piperno.
\newblock Filter Models for a Parallel and Non Deterministic 
  $\lambda$-calculus.
\newblock LNCS 711, pages 403-412, 1993.
  \bibitem[Gir87]{Gir87}
J.-Y. Girard.
\newblock Linear logic.
\newblock {\it Theoretical Computer Science}, 50:1--102, 1987.

\bibitem[GLT89]{GLT89}
J.-Y. Girard, Y. Lafont, and P. Taylor.
\newblock Proofs and Types.
\newblock Cambridge University Press, 1989.

\bibitem[Gir95a]{Gir95a}
J.-Y. Girard.
\newblock Geometry of Interaction III: accommodating the additives.
\newblock {\it Advances in Linear Logic, London Mathematical Society Lecture
  Notes Series} 222, 1995.

\bibitem[Gir95b]{Gir95b}
J.-Y. Girard.
\newblock Proof-nets: the parallel syntax for proof-theory.
\newblock In Ursini and Agliano, editors, 
\newblock {\it Logic and Algebra, New York, Marcel Dekker}, 1995.

\bibitem[Gir95c]{Gir95c}
J.-Y. Girard.
\newblock Light Linear Logic. 
\newblock Available by ftp anonymous on {\tt lmd.univ-mrs.fr}, in 
{\tt pub/girard}, 1995.

\bibitem[Gir96]{Gir96}
J.-Y. Girard.
\newblock Coherent Banach Spaces : Continuous Denotational Semantics 
(Extended Abstract).
\newblock {\it Electronic Notes in Theoretical Computer Science}, 3, 1996.

%\bibitem[Mat97]{Mat97}
%S.Matsuoka.
%\newblock Nondeterministic Linear Logic (Preliminary Results).
%\newblock  In Procs of Workshop of Programming Research Group of 
%Information Processing Sciety of Japan, 
%97-PRO-12, pages 21-26, 1997.

\bibitem[Pra95]{Pra95}
V.R. Pratt.
\newblock Chu Spaces and their Interpretation as Concurrent Objects.
\newblock LNCS 1000, pages 392-405, 1995.

\bibitem[Pra97]{Pra97}
V.R. Pratt.
\newblock Towards Full Completeness of the Linear Logic of Chu Spaces.
\newblock {\it Electronic Notes in Theoretical Computer Science}, 6, 1996.

\end{thebibliography}
\end{document}